\documentclass[a4paper,11pt]{article}
\usepackage{pos}
\usepackage{amsmath}
\usepackage{braket} 
\usepackage[utf8]{inputenc}

\title{Toward quantum computations of the $O(3)$ model using qumodes}

\author*[a]{Raghav G.~Jha}
\author[a,b]{Felix~Ringer}
\author[c]{George~Siopsis,}
\author[c]{Shane~Thompson}

\affiliation[a]{Thomas Jefferson National Accelerator Facility, Newport News, VA 23606, USA}
\affiliation[b]{Department of Physics, Old Dominion University, Norfolk, VA 23529, USA}
\affiliation[c]{Department of Physics and Astronomy, University of Tennessee, Knoxville, Tennessee 37996-1200, USA}

\FullConference{40th International Symposium on Lattice Field Theory at Fermilab \\
July 31- August 4, 2023}

\abstract{ 
We express the discrete 1+1-dimensional $O(3)$ non-linear sigma model (NL$\sigma$M) in a form well-suited for the continuous variable approach to quantum computing. Within the Schwinger boson formulation, we need two qumodes (quantum-mechanical oscillators) at each lattice site. We envision that it might be possible to reach the scaling regime of this model and observe asymptotic freedom on near-term photonic quantum devices in the coming decade.  
}
\begin{document}
\maketitle
\section{Introduction} 
It was already pertinent in the 1970s that for the computation of some of the physical problems especially involving quantum systems, the classical computer would take an exponentially large time to carry out a systematic computation. These problems were categorized within the NP complexity class. In order to approach these problems, it seemed inevitable that a paradigm shift was required. This led the community to think beyond computers that obeyed classical information theory. Following earlier works by Manin, Bennett, Toffoli, and others, a major breakthrough was provided by Feynman who remarked that quantum computers are the most natural way of simulating any quantum system. Since nature is quantum-mechanical at its most fundamental level, the computations should make use of this to carry out efficient calculations. This motivation has led to more than four decades of efforts to build and effectively simulate physical systems using computers based on the principles of quantum mechanics. Though, we are not close to the resources required to study say real-time evolution of continuum field theories or to understand the behavior of quantum systems even in lower dimensions yet, it is an active field of research. Irrespective of the available hardware resources, it is useful to think about methods of how one can study such systems if and when we have the required quantum hardware.

For lattice field theorists, gauge theories with interesting dynamics and properties are the obvious playgrounds to test the ideas and formulate them in a way amenable to quantum hardware. In the absence of a framework to directly deal with gauge theories in higher dimensions using quantum hardware, much work to date has been restricted to 1+1 dimensions. Even in these simpler cases, some useful models are available to test the ideas of dynamical mass gap, confinement, and asymptotic freedom. One such model is the $O(3)$ NL$\sigma$M. In this work, we will start with the well-known rotor formulation of this model and express it in terms of bosonic oscillators which are the building blocks of continuous-variable (CV) quantum computing. The theory of quantum information based on CVs which is encoded in harmonic oscillators, provides an alternative approach for quantum computing \cite{2005Brau,2012Weed} alongside the qubit (or digital) approach. They have also been shown to be useful toward the long-term goal of fault-tolerant quantum computing \cite{2001Gott}. 
One often refers to the idea of `qumodes' as similar to `qubits' and `qudits'. A qumode is a quantum-mechanical harmonic oscillator with infinitely discrete quantum states. It is these quantum states which we will use to encode our information, unlike the qubits where only $\vert 0 \rangle$ and $\vert 1 \rangle$ are used. In this conference paper, we discuss how to express the $O(3)$ model in the oscillator basis using the Schwinger boson approach and comment on the possibility of whether this is realistic to achieve using current state-of-the-art methods in photonic quantum hardware. The $O(3)$ model has been the subject of many investigations recently for efficient time evolution using fuzzy qubitization methods~\cite{Alexandru:2022son} and for preparation of ground states using cold atoms~\cite{Ciavarella:2022qdx}. In addition, the Schwinger boson approach has also been used in Ref.~\cite{Davoudi:2022xmb} to formulate quantum algorithms for $SU(2)$ gauge theory. 
\section{Mapping the rotor model onto the oscillator basis}
The 1+1-dimensional $O(3)$ model is described by the action:
\begin{equation}
    S = \frac{\beta}{2}\int dx dt~\partial_{\mu} \textbf{n} \cdot \partial^{\mu} \textbf{n}, 
\end{equation}
where $\textbf{n}$ is a 3-vector and $\beta = 1/g^{2}$. We use the standard terminology and refer to $\beta=0$ as the strong-coupling limit and $\beta \to \infty$ as the weak-coupling limit. For studying this model using quantum computation methods, one often uses the continuous-time limit of the model. This limit was shown to be related to a rotor Hamiltonian~\cite{Hamer:1978ew}
and is given by:
\begin{equation}
\label{eq:Ham}
    \widehat{H} = \frac{1}{2\beta} \sum_{j=1}^{N} \textbf{L}^{2}_{j} - \beta \sum_{\langle jk \rangle} \textbf{n}_{j} \cdot \textbf{n}_{k},
\end{equation}
where $j$ and $k$ are the nearest neighbor sites on a spatial lattice, $\textbf{n}$ is a unit 3-vector at site $j$ and $\textbf{L}$ is the angular momentum. The fields take values on the manifold $\mathbb{S}^2$ following the geometric constraint $\vec{n}.\vec{n} = 1$. In order to reduce the infinite degrees of freedom of this $\hat{H}$, we must choose an angular momentum cutoff, which we refer to as $l_{\text{max.}}$. With this choice, we have 
$\sum_{l=0}^{l_{\text{max.}}}(2l+1) = (l_{\text{max.}} + 1)^{2}$ and therefore $\hat{H}$ is given by a Hermitian matrix of size $(l_{\text{max.}}+1)^{2N}$. If we fix the length of the vector, it can be expressed in terms of angles as:
\begin{equation}
\vec{n} = (\sin\theta \cos \phi, \sin\theta \sin \phi, \cos\theta). 
\end{equation}
Note that this parametrization reduces the degrees of freedom from the outset because we restrict the length fluctuations. Another way to approach the $O(3)$ model is to consider the 3-vector as a collection of real scalar fields  with a potential that imposes the constraint and where the fluctuations are allowed but cost more (so they are not favored). Both these methods are equally promising pathways to simulating the NL$\sigma$M, however, it is not obvious as to which approach would be more optimal with respect to quantum resources. In order to set up the continuous variable approach to the quantum computation of this model, we have to express the Hamiltonian (\ref{eq:Ham}) in terms of a collection of oscillators. To do this, we note that the interaction term can be written as: 
\begin{align}
\label{eq:ni_nj}
    \textbf{n}_{j} \cdot \textbf{n}_{k} &= \sin(\theta_j)\sin(\theta_k) \cos(\phi_j - \phi_k) + \cos(\theta_j)\cos(\theta_k) \nonumber \\
    &= \cos(\theta_j)\cos(\theta_k) +
    \frac{1}{2} \Big[\sin(\theta_j)e^{i \phi_j} \sin(\theta_k)e^{-i \phi_k} + \text{H.c} \Big]. 
\end{align}
where H.c denotes hermitian conjugate. In order to rewrite the Hamiltonian in terms of oscillators, we note that the trigonometric functions appearing in (\ref{eq:ni_nj}) can be connected to the $\ket{lm}$ basis via the standard definitions of the spherical harmonics 
$Y_{lm}$ and the $\ket{lm}$ basis can be related to the combined Fock states of two harmonic oscillators ($a$ and $b$) given by $\ket{n_a, n_b}$ via the relation: 
\begin{equation}
\label{eq:sch_65} 
\vert l, m \rangle = \frac{(a^{\dagger})^{l+m} (b^{\dagger})^{l-m}}{\sqrt{(l+m)! (l-m)!}} \vert 0,0 \rangle.
\end{equation}
Here we have written two kinds of Bose operators $a$ and $b$ to invoke the relationship between the two-dimensional harmonic oscillator 
and the angular momentum basis with $[a,b] = [a^{\dagger}, b^{\dagger}] = 0$ and $[a, a^{\dagger}] = [b, b^{\dagger}] = 1$.
This representation is referred to as the Jordan map or Jordan-Schwinger representation. Note that we have suppressed the hats over $a$ and $b$ operators. The total number of bosons is $n = n_{a} + n_{b}$. We can define the three operators, which generate the Lie algebra $\mathfrak{su}(1,1)$ given by: 
\begin{equation}
\label{eq:opsSU11}
    K_{+} = a^{\dagger} b^{\dagger}, ~K_{-} = ba, ~K_{3} = \frac{1}{2} \Big(a^{\dagger}a + b^{\dagger}b + 1\Big). 
\end{equation}
Using the relation between the trigonometric function to spherical harmonics: 
\begin{align}
\label{eq:RR1}
   \cos(\theta) Y_{l,m}(\theta, \phi) &= \sqrt{\frac{(l-m+1)(l+m+1)}{(2l+1)(2l+3)}} 
   Y_{l+1,m}(\theta, \phi) + 
   \sqrt{\frac{(l-m)(l+m)}{(2l-1)(2l+1)}} 
   Y_{l-1,m}(\theta, \phi), 
\end{align}
and (\ref{eq:sch_65}), we find that the $\cos(\theta_j)$ term in the 
$O(3)$ Hamiltonian is of the form: 
\begin{align}
\label{eq:costerm} 
\cos(\theta_j) &= \frac{1}{\sqrt{(n_j + 1)(n_j - 1)}} (a^{\dagger}_j b^{\dagger}_j)  + \frac{1}{\sqrt{(n_j + 1)(n_j + 3)}} (b_{j} a_{j}) \nonumber \\
&= \frac{1}{\sqrt{(n_j + 1)(n_j - 1)}} K_{+}  + \frac{1}{\sqrt{(n_j + 1)(n_j + 3)}} K_{-}, 
\end{align}
while the other terms are: 
\begin{equation}
\label{eq:sin_iphi} 
    \sin(\theta_j) e^{i \phi_j} = \Big[\frac{1}{\sqrt{(n_j+1)(n_j+3)}} (b_{j}b_{j}) -\frac{1}{\sqrt{(n_j+1)(n_j-1)}} (a^{\dagger}_{j}a^{\dagger}_{j})\Big],
\end{equation}
and its hermitian conjugate. We have used the map between oscillator occupation number and angular momentum basis: $l = (n_{+} + n_{-})/2$ and $m = (n_{+} - n_{-})/2$ such that $n = 2l, l+m = n_{+}, l-m = n_{-}$. 
We also note that if we define $J_{+} = a^{\dagger}b$ which is one of the operators that form the representation of $\mathfrak{su}(2)$, then $[K_+, J_+] = -a^{\dagger}a^{\dagger}$ and $[K_-, J_+] = bb$
and (\ref{eq:sin_iphi}) can be written in terms of these operators. In order to obtain (\ref{eq:costerm}), we made use of $(b^{\dagger})^{l-m} \vert 0,0 \rangle = \sqrt{(l-m)!} \vert 0,n_{-} \rangle$
and $(a^{\dagger})^{l+m} \vert 0,0 \rangle = \sqrt{(l+m)!} \vert n_{+},0 \rangle$. 
The action of various operators is summarized below:
\begin{align}
    K_{-} \ket{lm} & = \sqrt{(l-m)(l+m)} \ket{l-1,m}  \label{eq:ops1}  \\
    K_{+} \ket{lm} & = \sqrt{(l+m+1)(l-m+1)} \ket{l+1,m} \label{eq:ops2}  \\
    J_{+} \ket{lm} & = \sqrt{(l-m)(l+m+1)} \ket{l,m+1} \label{eq:ops3}. 
\end{align}
Using (\ref{eq:ops1}) and (\ref{eq:ops3}), we can compute the action of $[K_-, J_+] = bb$ on the eigenket as: 

\begin{align}
[K_-, J_+]\ket{lm} &= (K_{-}J_{+} - J_{+}K_{-}) \ket{lm} \nonumber \\ 
& = \Big((l+m+1) \sqrt{(l-m)(l-m-1)} - 
(l+m) \sqrt{(l-m)(l-m-1)}\Big) \ket{l-1,m+1}  \\
& = \sqrt{(l-m)(l-m-1)} \ket{l-1,m+1} \,.
\end{align}
We then find that the kinetic term is given by:  
\begin{equation}
\label{eq:kinterm}
    \textbf{L}_{j} \cdot \textbf{L}_{j} = \frac{n_{j}}{2} \Big(\frac{ n_{j}}{2} + 1\Big),
\end{equation}
where $n_{j}$ is the number operator at site $j$. 
The maximum number of allowed bosons at each site for a particular choice of truncation is given by $2l_{\text{max.}}$ and each finite-dimensional operator in (\ref{eq:opsSU11}) is of size 
$(l_{\text{max.}}+1)^2$ and the size of the Hamiltonian for $N$ sites is $(l_{\text{max.}}+1)^{2N}$ similar to the size of the rotor Hamiltonian in (\ref{eq:Ham}). Using (\ref{eq:ni_nj}), (\ref{eq:costerm}), (\ref{eq:sin_iphi}), its hermitian conjugate and (\ref{eq:kinterm})  in (\ref{eq:Ham}) we can write the Hamiltonian \emph{entirely in terms of Bose operators}. 

The suitable choice of truncation required depends strongly on the coupling. For $\beta \to 0$, a hard truncation to the smallest angular momentum states works well. However, as
$\beta \sim O(1)$, one needs to consider larger values of $l_{\text{max.}}$. From the extensive study of this model using classical methods, it is known that the scaling regime for the continuous-time model is around $\beta \sim 1.2$ if $N$ is taken sufficiently large. We will show in the next section using exact diagonalization methods that $l_{\text{max.}} \sim 4$ appears to be sufficient for this model and this is corroborated by the state-of-the-art tensor network results. This translates to  8 bosons at each site. The current photonic hardware methods have been able to study about 15 photons i.e., $\ket{n \sim 15}$~\cite{Johnson2010, 2020PhRvX..10b1060W}. So, in principle, the truncation over the states is already possible with current technology on a few sites. It remains to be seen whether this can be achieved in practice. 
\section{Exact Diagonalization and beyond}
\begin{figure}
	\centering 
	\includegraphics[width=0.65\textwidth]{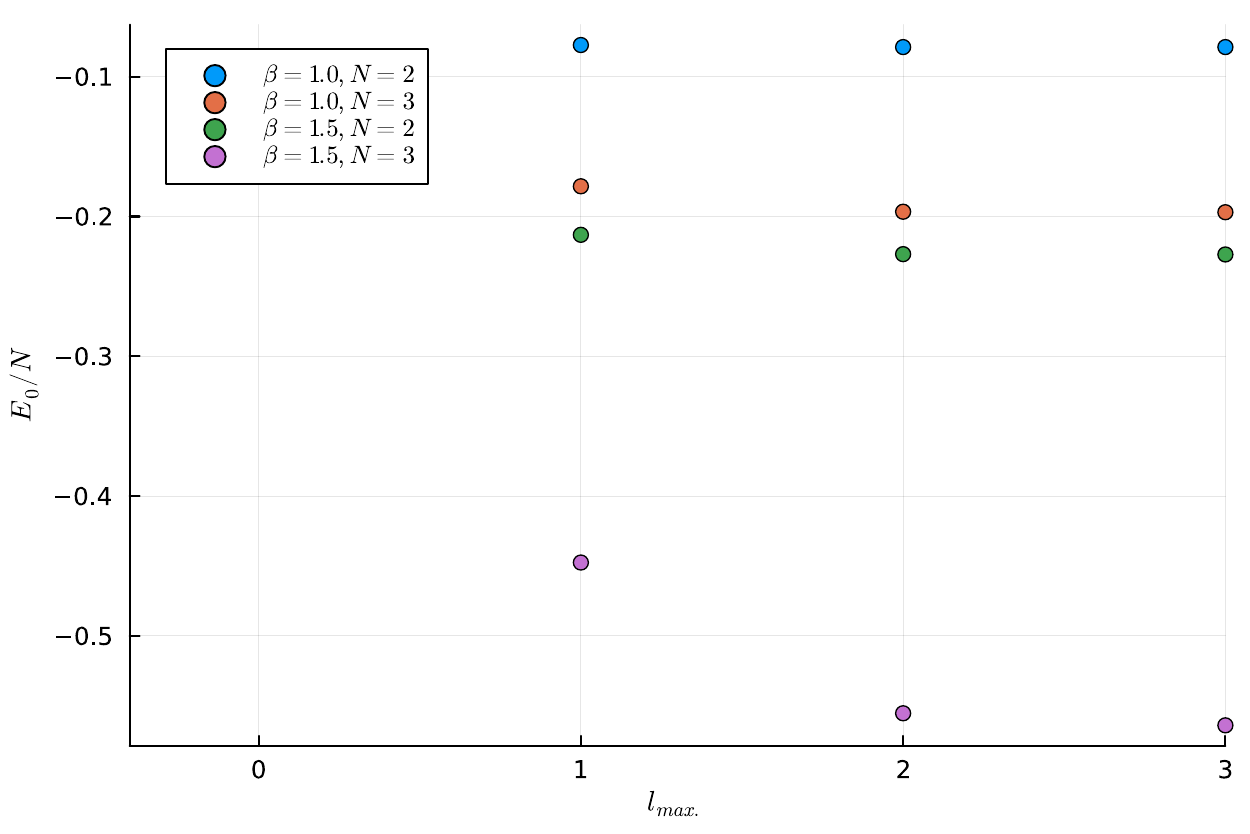}	\caption{\label{fig:ED1}The ground state energy density of the $O(3)$ model for $N=2,3$ with different truncation. We see that the ground state energy density decreases as we increase $N$ and it appears to converge beyond $l_{\text{max.}} = 3$. }
\end{figure}

\begin{figure}
	\centering 
	\includegraphics[width=0.65\textwidth]{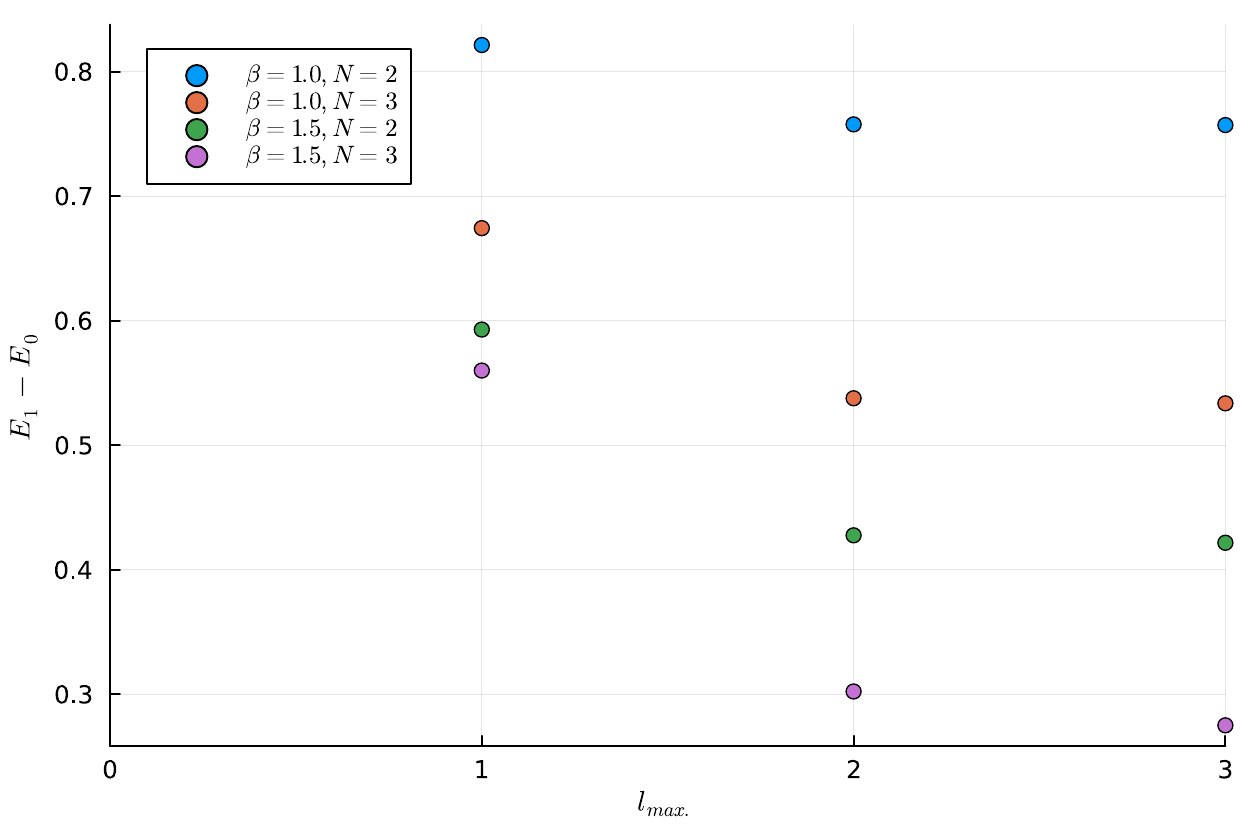}\caption{\label{fig:ED2}The gap computed using the ED method. The gap seems to decrease with including higher angular momentum states for the couplings considered.} 
\end{figure}

For a small number of lattice sites and modest truncation, we can compute the energy gap and ground state energy using the exact diagonalization (ED) method. The state-of-the-art classical method is tensor networks and this model has been studied using matrix product states (MPS) with and without the topological $\theta$-term \cite{Bruckmann:2018usp, Tang:2021uge}
We show the ED results in Fig.~\ref{fig:ED1} and Fig.~\ref{fig:ED2} respectively. This result has been accurately computed from tensor network computations (MPS) with $N=40$ (see Fig.~1 of Ref.~\cite{Bruckmann:2018usp}). In order to realize this model on hardware in the coming decades, it is crucial that we understand the scaling window and the effects of truncation. If we had to strictly take the $\beta \to \infty$ limit for the continuum limit, then it would likely be hopeless \footnote{In the limit of large $\beta$, the RG flow to the UV fixed point might make some parameters irrelevant and there might be simplification of the Hamiltonian. In such case, the implementation on cavity QED like hardware might be easier. We leave this analysis for future work.} to realize this on photonic quantum hardware due to the large occupation number states involved. However, earlier studies have concluded that $\beta \sim 1.2$ is sufficient to extract the interesting Physics. 
In order to understand this, it is useful to consider the mass gap as a function of $\beta$. The result is known from Ref.~\cite{Hasenfratz1990, Bruckmann:2018usp} to be $m = 128\pi \beta \exp(-2\pi\beta)$ as $\beta \gg 1$.  
The large coefficient ensures that the continuum limit can be reached for a modest value of $\beta$ because the theory remains in the same phase beyond some $\beta \sim \mathcal{O}(1)$. We believe that this would result in the near-term realization of asymptotic safety in the $O(3)$ model.  

Once we have the Hamiltonian in terms of oscillators with some fixed truncation of the Fock space, we can think about implementing the unitary operator $\exp(-iHt)$. The Hamiltonian consists of nearest-neighbor interactions, i.e. it is 4-local in qumodes and is sparse. For example, the rotor Hamiltonian of the model is $7$-sparse\footnote{A matrix is $d$-sparse if at most $d$ elements are non-zero in a given row or column} for $N=2$ and $l_{\text{max.}} = 2$. It is known that Hamiltonian simulation methods are efficient for 1-sparse matrices since the error grows at most like $\mathcal{O}(d)$. So, the first step is to write $H = \sum_{i=1}^{d} H_{d}$ as a sum of 1-sparse matrices and then express the time evolution operators as linear combination of unitaries (LCU).  
An additional complication is the fact that the Hamiltonian has coefficients depending on the inverse square roots of the occupation number operator. These floating-point operations can be done using an adder, Newton's method of finding square roots, and multiplier circuits by coupling the photonic hardware to qubits (hybrid systems). It can also be implemented using an equivalent continuous variable (CV) gate set. We leave the detailed implementation of the time evolution using the Schwinger boson formalism for future work. 
\section{Conclusion}
In this work, we considered the $O(3)$ model, which has several features similar to QCD. Due to its reduced dimensionality, it provides a suitable arena to test the applications of quantum hardware in the coming decades. Though this model is not a gauge theory like the Schwinger model, it is difficult to implement due to the larger continuous global symmetry group. Rather than studying the model using the qubit-based approach, we provided a construction that is well-suited for the continuous-variable approach to quantum computation. This was achieved by expressing the $O(3)$ model in terms of two oscillators (qumodes) at each site. Our construction does not assume any truncation over angular momentum states at the start and, hence, it is more general than other approaches. Though, we do have to impose a cut-off in the Fock basis for practical calculations. The truncation effects become severe at weak couplings ($\beta \gg 1$) closer to the continuum limit. However, the established results from classical computations show that the onset of the scaling limit is well captured with the truncation imposed at rather small values of $\beta$. This signifies that this model can provide a realistic scenario for accessing physical lattice length scales using quantum computing methods. The truncation also sets the maximum number of bosons (photons) at each site (cavity) needed for the experimental realization. Based on our rough estimate, it appears that it will be possible to study the model on a few sites using photonic hardware and a suitable gate set in the coming decade. This would imply that we can potentially observe asymptotic freedom on near-term quantum devices for the first time in this model. 
\vspace{10 pt}

\noindent {\textbf{Acknowledgments:}} 
RGJ would like to thank Robert Edwards, Steve Girvin, Kostas Orginos, and Nathan Wiebe for discussions. RGJ was supported by the U.S. Department of Energy, Office of Science, National Quantum Information Science Research Centers, Co-design Center for Quantum Advantage under contract number DE-SC0012704. RGJ and FR are supported by the DOE with contract No. DE-AC05-06OR23177, under which Jefferson Science Associates, LLC operates Jefferson Lab. GS and ST acknowledge support by DOE ASCR funding under the Quantum Computing Application Teams Program, the Army Research Office award W911NF-19-1-0397, NSF award DGE-2152168, and DOE award DE-SC0023687. 

\bibliographystyle{JHEP}
\bibliography{lattice23.bib}
\end{document}